\documentclass{PoS}
\usepackage{url}
\usepackage{subfigure}

\usepackage{cite}

\title{Charged Higgs production via vector-boson fusion at NNLO in QCD}

\ShortTitle{Charged Higgs production via vector-boson fusion at NNLO in QCD}

\author{\speaker{Marco Zaro}\\
   Universit\'e Catholique de Louvain - CP3\\
       E-mail: \email{marco.zaro@uclouvain.be}}

\author{Paolo Bolzoni\\
Institut f\"ur Theoretische Physik, Universit\"at Hamburg\\
       E-mail: \email{paolo.bolzoni@desy.de}}

\author{Fabio Maltoni\\
       Universit\'e Catholique de Louvain - CP3\\
       E-mail: \email{fabio.maltoni@uclouvain.be}}
       
       \author{Sven-Olaf Moch\\
       Deutsches Elektronen-Synchrotron, DESY\\
       E-mail: \email{sven-olaf.moch@desy.de}}

\abstract{We present the total cross sections at next-to-next-to-leading order (NNLO) in the strong coupling for single and double charged Higgs production via weak boson fusion. Results are obtained via the structure function approach, which builds upon the approximate, though very accurate, factorization of the QCD corrections between the two quark lines. The theoretical uncertainty on the total cross sections at the LHC from higher order corrections and the parton distribution uncertainties are estimated at the 2\% level each for a wide range of Higgs boson masses.}

\FullConference{Third International Workshop on Prospects for Charged Higgs Discovery at Colliders - CHARGED2010,\\
		September 27-30, 2010\\
		Uppsala Sweden}

\begin{document}

\section{Introduction}
After almost six months of running at 7 TeV and about 40 inverse picobarns of collected integrated luminosity, the LHC collider and its experiments are showing excellent performances. Both strong and electroweak interactions seem to be properly understood, and therefore the experiments are ready to discover and study new physics. Even if the Higgs boson will be discovered as it is predicted by the Standard Model (SM), an extension will be somehow needed, in order to solve other still-open problems, like the hierarchy problem or the mechanism that gives mass to neutrinos. In many extension of the SM also the Higgs sector is expanded by adding another $SU(2)$ doublet or triplet, giving rise to several neutral Higgs bosons, as well as to some charged ones.\\
\section{Charged Higgs boson production}
\begin{figure}[h!]
\subfigure[]{
\includegraphics[width=0.3\textwidth]{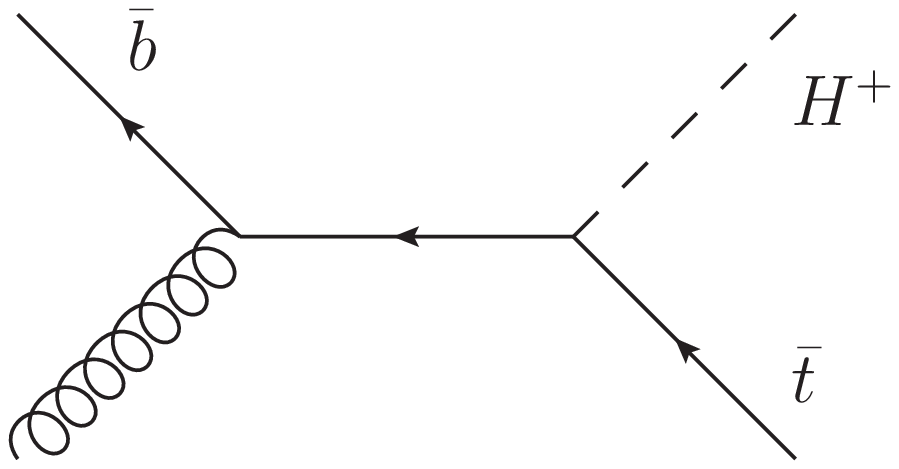}
\label{bght}
}\hspace{2cm}
\subfigure[]{
\includegraphics[width=0.3\textwidth]{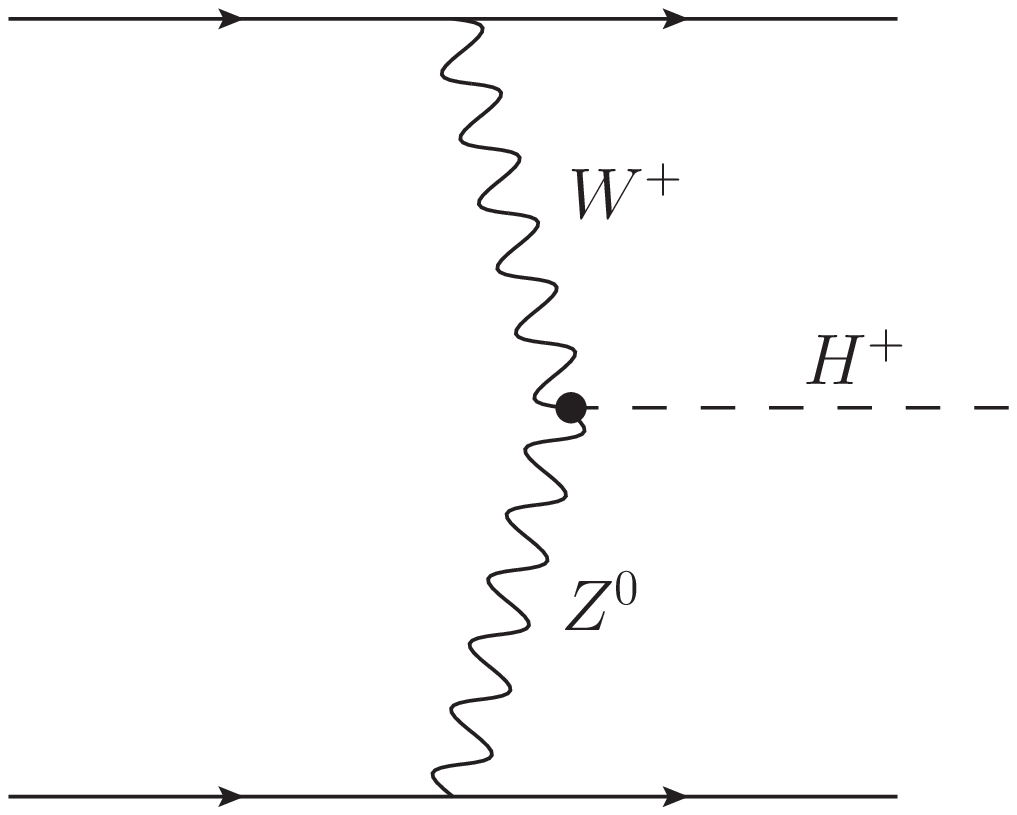}
\label{wbf}
}
\caption{(a): One of the diagrams for the associated production of a charged Higgs and a $t$-quark. 
(b): The production of a charged Higgs via VBF.}
\end{figure}

The production channels for charged Higgses strongly depend on the model considered. In models where the Higgs is embedded in isospin doublets (2HDM, MSSM, ...), the dominant production channel is the associated production with a top quark in bottom-gluon fusion~\cite{Berger:2003sm, Weydert:2009vr} (see Fig.~\ref{bght}). In such models isospin symmetry forbids the $W^\pm H^\mp Z$ and $W^\pm H^\mp \gamma$ vertices at tree level, while they can be loop-induced~\cite{Kanemura:1997ej}. For this reason, both the vector boson fusion (VBF) and the Higgs-strahlung processes have a negligible cross-section in these models.\\
This situation changes if we consider models with isospin triplets in the Higgs-sectors (see~\cite{Godfrey:2010qb} and refs. therein). Many versions of such models exist in literature, with a very interesting particle content, including doubly charged Higgses. Moreover, these models allow the coupling of the charged Higgs(es) to gauge bosons, with a strength proportional to the ratio $\frac{v'}{v}$, where $v'$ is the triplet vacuum expectation value (vev) and $v$ the EW one. Usually $v'$ is constrained by electroweak data (in particular by the SM $\rho$ parameter) to be at most of a few GeV, but this bound can be loosened by including more triplets in the model~\cite{Georgi:1985nv}. If this is the case, the production of singly and doubly charged Higgs bosons via VBF can be studied at colliders~\cite{Gunion:1989ci, Vega:1989tt, Asakawa:2006gm}.\\
The VBF (see Fig.~\ref{wbf}) is one of the most interesting production channels for the SM Higgs boson, because of the total cross-section size (about 1.3~pb for $m_h=120$~GeV at the LHC with $\sqrt s =7$~TeV) and the clear experimental signature. The Higgs boson is mainly produced at central rapidities, while the outgoing quarks might be detected as hard jets in the forward/backward directions. The VBF being an electroweak process without color exchange between the two protons at tree level, extra hadronic activity is mainly outside the central rapidity region. This allows the Higgs boson decay to be detected and studied. \\
If the bounds coming from the $\rho$ parameter can be evaded, these features remain true also for charged Higgses. 
In addition to this, the SM Higgs production via VBF is one of the best theoretically known processes for the LHC: fully differential codes exist for the NLO QCD correcitons~\cite{Nason:2009ai} as well as for the combined EW + QCD NLO corrections~\cite{Ciccolini:2007ec}, and the total cross-section can be computed up to NNLO in QCD~\cite{Bolzoni:2010xr, Bolzoni:2010as} exploiting the (almost exact) so-called structure function approach~\cite{Han:1992hr}.
\section{The structure function approach to VBF}
The structure function approach consists in considering VBF as a double deep-inelastic scattering (DIS), relying on the fact that already at LO the interferences between different diagrams are kinematically suppressed at the per-mil level, and that at NLO the QCD corrections involving the exchange of a gluon between the protons vanish because of color conservation. Hence, from the knowledge of the QCD corrections to the DIS structure functions~\cite{Bardeen:1978yd, Kazakov:1990fu, Zijlstra:1992kj, Zijlstra:1992qd, Moch:1999eb}, it is possible to compute the total VBF cross-section.\\
This kind of factorization, which is exact at NLO in QCD, apparently fails for the NNLO corrections. The color conservation argument cannot be used anymore for diagrams with a double gluon exchange between protons, and also diagrams involving a $t-b$-quark loop have to be taken into account. However the impact of these two kinds of diagrams on the total cross-section is estimated to be indeed negligible if compared to the factorizable corrections.\\
With very little effort this approach can be extended to the VBF production of charged Higgses, thus allowing for the prediction of the total cross-section up to NNLO in QCD. The Feynman rules for a wide class of models with triplets are listed in~\cite{Godfrey:2010qb}. 

\section{Results}
We now show some results for the total cross section of the production of a charged Higgs boson via VBF. In our computation we assume a $VVH$ vertex of the form
\begin{equation}
\Gamma ^{\mu\nu}_{V_iV_jH}=2 \left(\sqrt 2 G_F\right)^{1/2} m_i m_j \; F_{i j}\left(-i g^{\mu\nu}\right)
\end{equation}
where the dimensionless coefficients $F_{ij}$ depend on the particular model. In order not to loose generality, we will show cross-sections computed with $|F_{ij}|=1$.\\
\begin{figure}[h!]
\subfigure[]{
\includegraphics[width=0.45\textwidth]{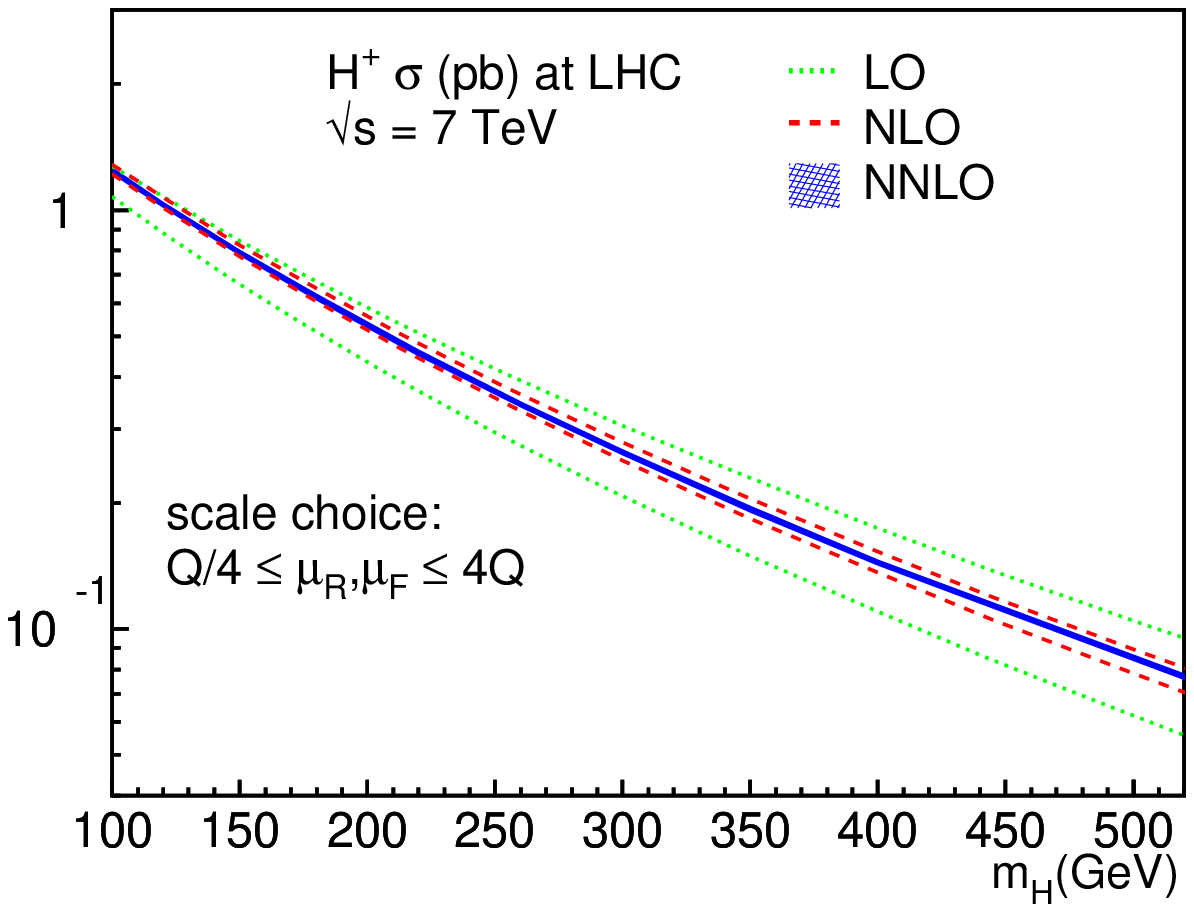}
}\hspace{0.5cm}
\subfigure[]{
\includegraphics[width=0.45\textwidth]{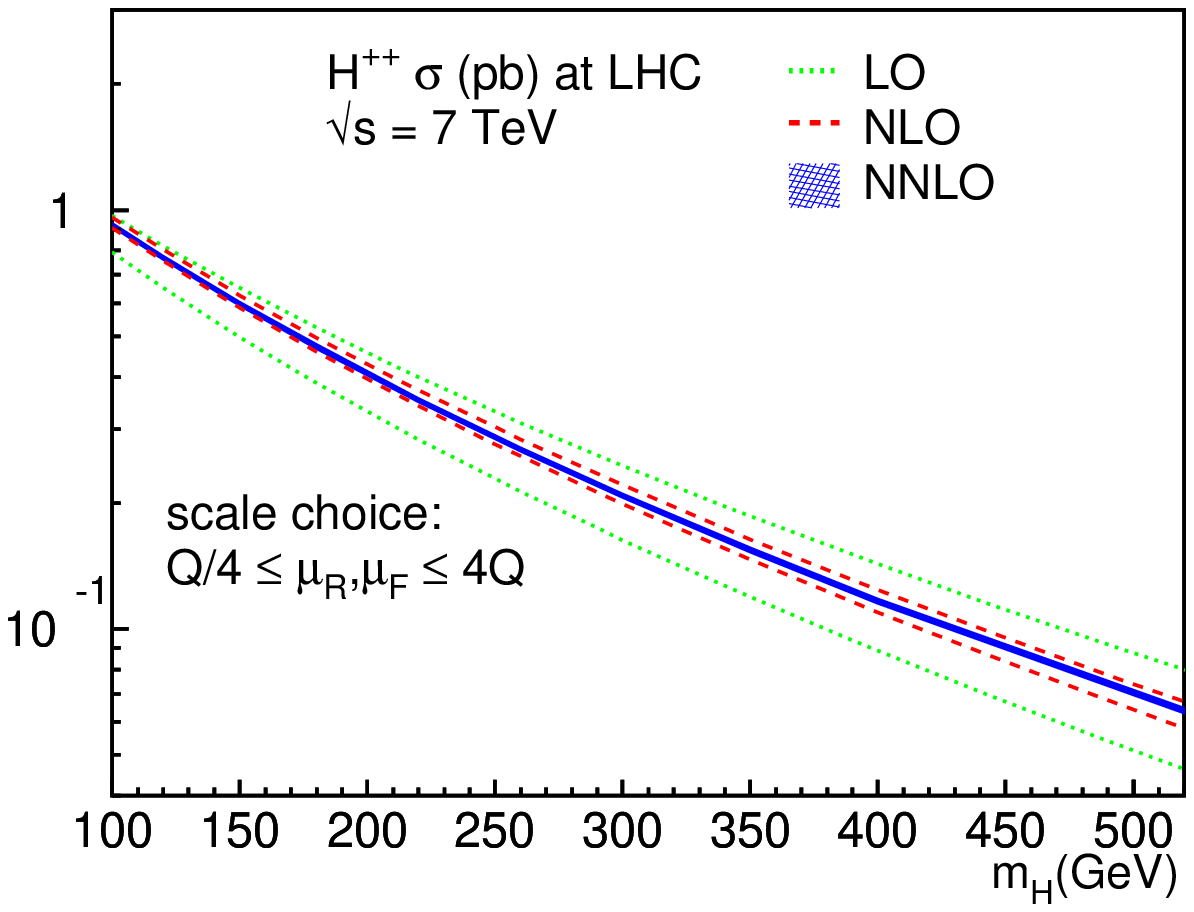}
}
\caption{(a): The total cross-section for the singly charged Higgs production via VBF at the LHC, with $\sqrt s = 7$TeV. The MSTW 2008~\cite{Martin:2009iq} PDF set has been used. The uncertainity bands are obtained from the variation of the renormalization and factorization scale in the interval $1/4 Q<\mu_f,\mu_r<4 Q$, where $Q$ is the virtuality of the vector boson.
(b): Same as (a) for a doubly charged Higgs. }
\label{hctot}
\end{figure}

\begin{figure}[h!]
\subfigure[]{
\includegraphics[width=0.45\textwidth]{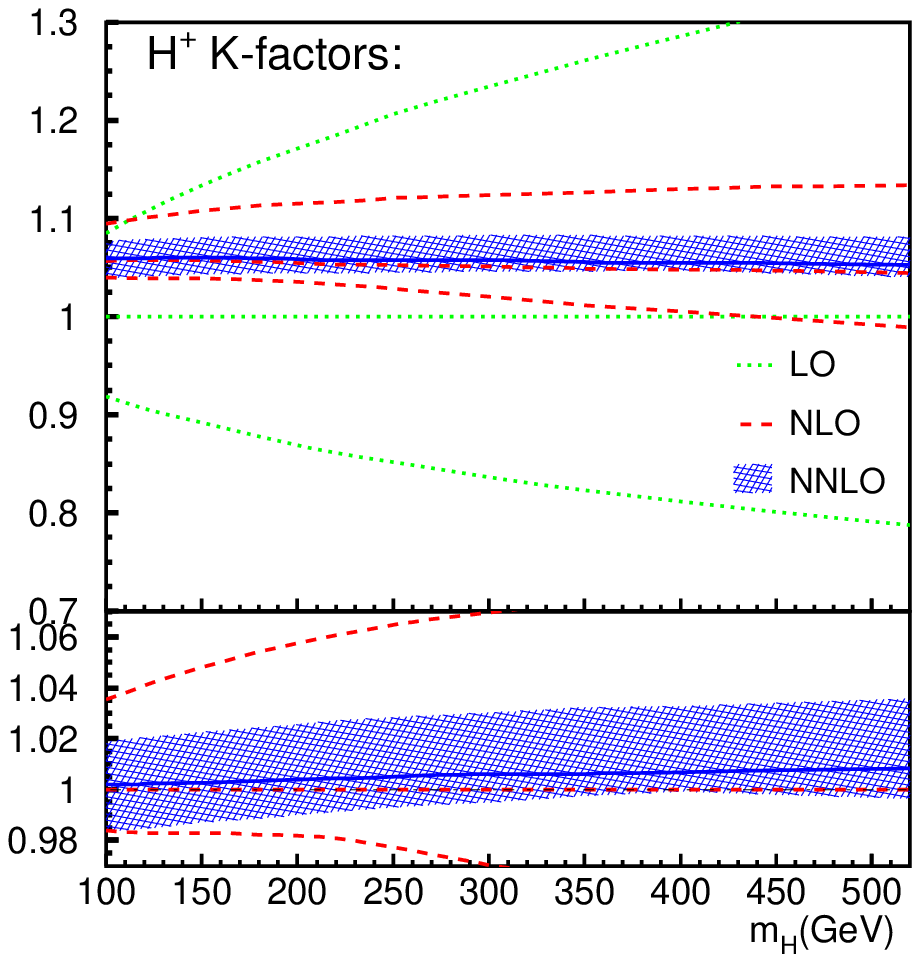}
}\hspace{0.5cm}
\subfigure[]{
\includegraphics[width=0.45\textwidth]{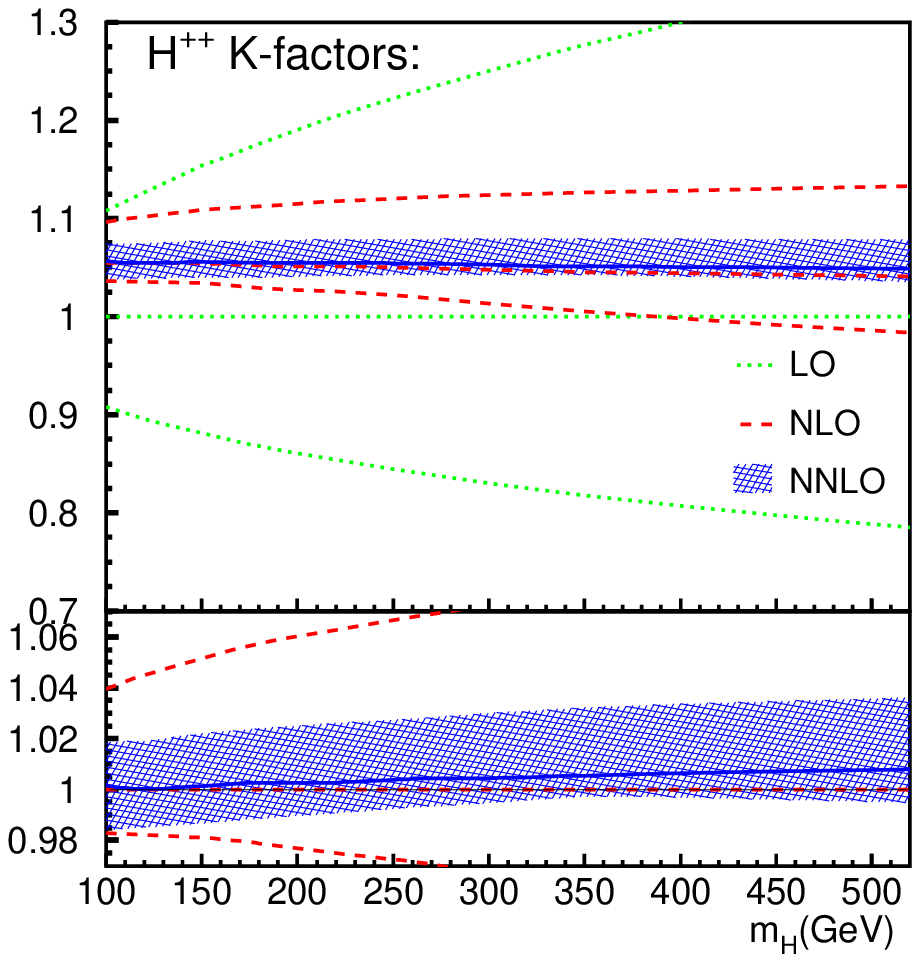}
}
\caption{(a): The NLO/LO (upper inlay) and NNLO/NLO (lower inlay) $K$-factors for sinlgy charged Higgs production via VBF at the LHC.
(b): Same as (a) for a doubly charged Higgs. }
\label{hck}
\end{figure}

\begin{figure}[h!]
\subfigure[]{
\includegraphics[width=0.45\textwidth]{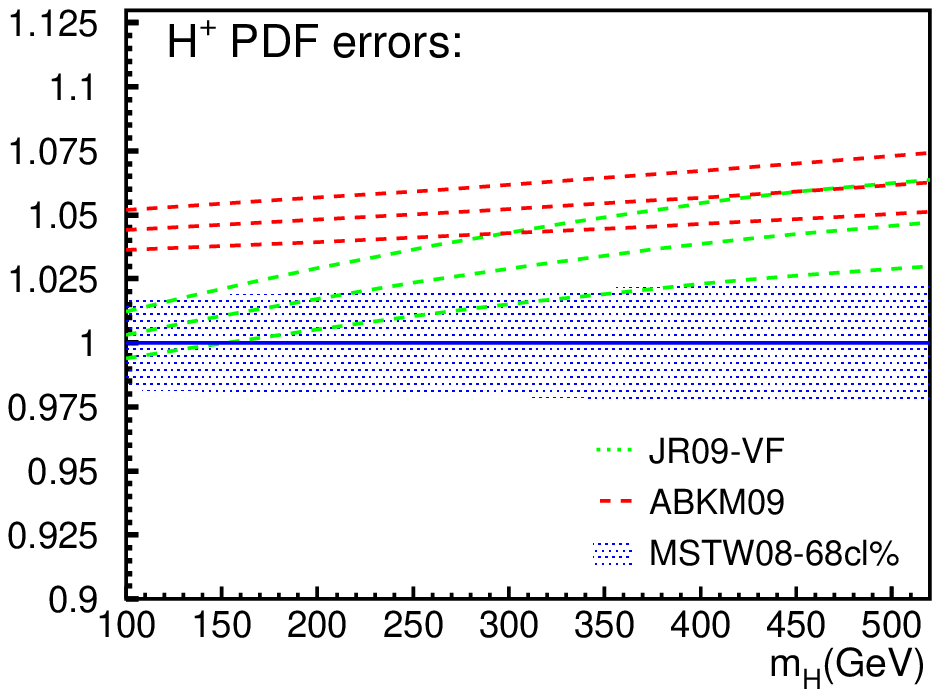}
}\hspace{0.5cm}
\subfigure[]{
\includegraphics[width=0.45\textwidth]{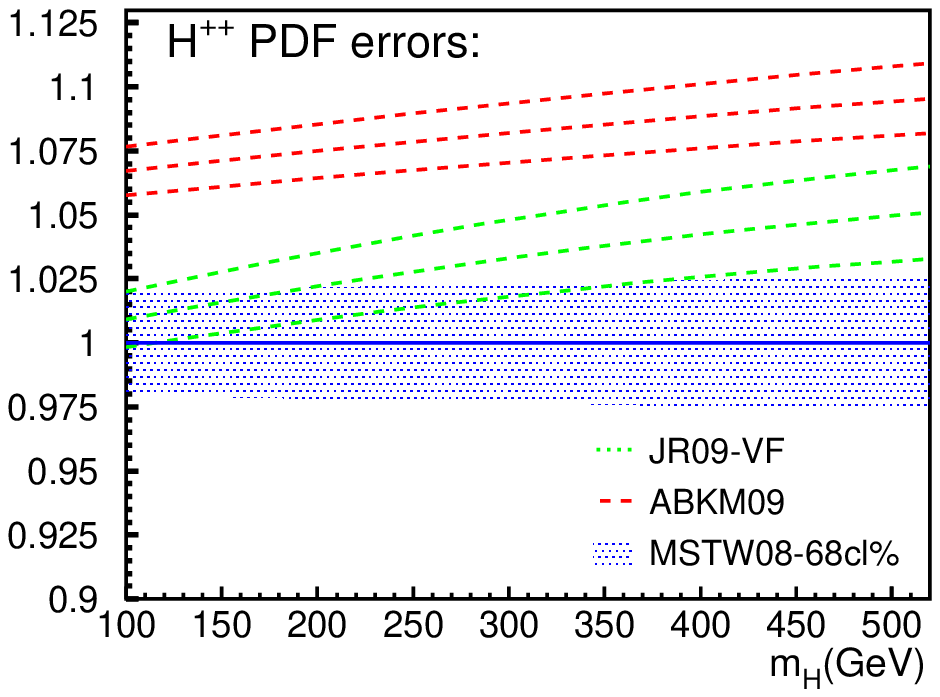}
}
\caption{(a): The PDF uncertainities for the total NNLO cross-section for the singly charged Higgs production via VBF at the LHC. We show the MSTW 2008~\cite{Martin:2009iq} 68\% (blue), the ABKM 2009~\cite{Alekhin:2009ni} (red) and the JR 2009~\cite{JimenezDelgado:2009tv} (green) sets, normalized to the cross-section obtained with the best fit PDF from MSTW.
(b): Same as (a) for a doubly charged Higgs. }
\label{hcpdf}
\end{figure}

In Fig.~\ref{hctot} we plot the total cross-section for single and double charged Higgs at the LHC ($\sqrt s= 7$TeV) via VBF. The theoretical uncertainity band is obtained by varying renormalization and factorization scale in the interval $1/4 Q<\mu_f,\mu_r<4 Q$, with $Q$ the virtuality of the vector boson. From those plots we can appreciate the convergence of the perturbative series and the improvement of the theoretical uncertainities. The scale uncertainities reduce to the 2\% level for the NNLO total cross-section, as illustrated by the $K$-factors shown in Fig.~\ref{hck}. In particular the impact of the NNLO corrections on the total cross-section is very small, hardly reaching the 1\% at large values of the Higgs mass. The LO, NLO and NNLO total cross-sections are computed using the corresponding parton distribution function (PDF) from the MSTW 2008~\cite{Martin:2009iq} set.\\
Finally, in order to estimate the theoretical uncertainities coming from PDFs, in Fig.~\ref{hcpdf} we compare the numbers obtained with the MSTW set with two others NNLO PDF sets available: the sets ABKM 2009~\cite{Alekhin:2009ni} and JR 2009~\cite{JimenezDelgado:2009tv} (in the variable flavour number scheme). From these plots we can see the need to consider more than one single PDF set to correctly estimate the PDF uncertainities, because the error bands from different sets do not overlap in a wide range of Higgs masses. The deviations from the center of the PDF ''envelope`` are of about $\pm 3\%$ and $\pm 5 \%$ for the singly and the doubly carged Higgs boson respectively.\\

\section{Conclusions}
We extended the computation done in~\cite{Bolzoni:2010xr} to the case of charged Higgs bosons. We computed the radiative QCD corrections to the total cross-section of charged VBF up to NNLO, showing the quality of the perturbative series convergence and the reduction of the theoretical uncertainities. The details of the computation will be soon presented in~\cite{Bolzoni:2011}. The total cross-sections for neutral (SM) and charged Higgses can be computed up to NNLO in QCD using the web interface~\cite{webint} for our code.

\end{document}